\documentclass[fleqn,12pt,twoside]{article}
\usepackage{espcrc1}

\usepackage{graphicx}
\usepackage[figuresright]{rotating}

\newcommand{\AmS}{{\protect\the\textfont2
  A\kern-.1667em\lower.5ex\hbox{M}\kern-.125emS}}

\title{Gauge invariance of the color-superconducting gap on the mass shell}

\author{R.D. Pisarski\address{Department of Physics,
                              Brookhaven National Laboratory, \\
                              Upton, NY 11973, U.S.A.}
        and
        D.H. Rischke\address{Institut f\"ur Theoretische Physik,
                             Johann Wolfgang Goethe--Universit\"at, \\
                             Robert-Mayer-Str.\ 10, 
                             D--60054 Frankfurt am Main, Germany}}
       
\begin{document}

\maketitle

\begin{abstract}
The gap parameter for color superconductivity is
expected to be a gauge invariant quantity, at least
on the appropriate mass shell. Computing the gap to 
subleading order in the QCD coupling constant, $g$, we show that
the prefactor of the exponential in $1/g$ is gauge dependent off 
the mass shell, and independent of gauge on the mass shell.
\end{abstract}

\section{INTRODUCTION}

Quantum chromodynamics (QCD) is the fundamental theory of the strong
interactions. In strongly interacting matter at large density or, 
equivalently, large quark chemical potential $\mu$, 
asymptotic freedom \cite{asymp} tells us that the QCD coupling
$g$, evaluated at the scale $\mu$, is logarithmically small, $g(\mu) \ll 1$.
Then single-gluon exchange, which is
attractive in the color-antitriplet 
channel \cite{bai84}, is the dominant interaction between two quarks
at the edge of the Fermi sea.  By Cooper's theorem \cite{BCS}, 
any attractive interaction destabilizes the Fermi surface and, 
at sufficiently small temperature $T$, leads to the condensation of 
Cooper pairs. In QCD, quark-quark Cooper pairs break the color
symmetry.  In this color superconductor, at $T=0$
it costs at least $2\phi_0$ to excite a particle-hole pair,
where $\phi_0$ is the value of the superconducting
gap function at the Fermi surface.
This gap $\phi_0$ can be computed using a mean-field approximation to a
self-consistent gap equation
with single-gluon exchange \cite{SchaferWilczek,pis00}.

Schematically, this gap equation can be written 
in the form \cite{pis00}
\begin{equation}
\phi^+_k = g^2 \, \int \frac{{\rm d}q}{\epsilon_q^+} \; \phi^+_q
\left[\; \zeta \, \ln \left( 
\frac{\mu^2}{|{\epsilon_k^+}^2-{\epsilon_q^+}^2|} \right) 
+  \beta  +  \beta' \, \epsilon_q^+ \, \ln \left( \frac{\mu}{\epsilon_q^+} 
\right) + \alpha\, \epsilon_q^+ \; \right] \;.
\label{gapequation}
\end{equation}
Here, $\phi^+_k$ is the gap function for quasiparticles,
taken on the quasiparticle mass shell.
In general, the gap function $\phi^+(K)$ depends on four-momentum 
$K^\mu \equiv (k_0,{\bf k})$. The quasiparticle mass shell is
defined by
\begin{equation}
k_0 = \epsilon_k^+\;\; , \;\;\;
\epsilon_k^+ \equiv \sqrt{(\mu-k)^2 +|\phi_k^+|^2}\; ,
\end{equation}
where $\epsilon_k^+$ is the quasiparticle excitation energy.
The gap function on the quasiparticle mass shell depends on
$k$ only, $\phi^+_k \equiv \phi^+(\epsilon_k^+,{\bf k})$.
At the Fermi surface, $\phi^+_\mu \equiv \phi_0$.
The integration variable $q$ in Eq.\ (\ref{gapequation}) 
is the kinetic energy of ultrarelativistic particles.
The prefactors $\zeta, \beta, \beta',$ and $\alpha$ 
are either constants or depend at most logarithmically on $g$.
In weak coupling, $g \ll 1$, the solution of Eq.\ (\ref{gapequation}) at the
Fermi surface is \cite{SchaferWilczek,pis00,son99,miransky}
\begin{equation}
\phi_0 = 2 \, b \, \mu \, \exp \left( - \frac{c}{g} \right) 
\left[ 1 + O(g) \right]\;.
\label{gapsolution}
\end{equation}
This solution implies $\ln (\mu/\phi_0) \sim 1/g$, such
that the various terms in Eq.\ (\ref{gapequation}) differ by powers of $g$. 
The first term $\sim \zeta$ contains two powers of the
logarithm $\ln (\mu/\phi_0)$. One is the well-known 
BCS logarithm \cite{BCS} which arises from the integration over 
$q$, 
\begin{equation}
\int \frac{{\rm d}q}{\epsilon_q^+} \sim 
\ln \left( \frac{\mu}{\phi_0} \right)\,\, .
\label{BCSlog}
\end{equation}
The other one is the logarithm multiplying
$\zeta$, when the momentum $q$ is near the Fermi
surface and $\epsilon_q^+ \sim \epsilon_k^+ \sim \phi_0$.
This logarithm is special to
theories with long-range interactions, like 
the exchange of almost static magnetic gluons in QCD  
\cite{pis00,son99}. Its origin is a collinear singularity when
integrating over the angle between incoming and outgoing quark
momenta in the gap equation. 
Since $\ln^2 (\mu/\phi_0) \sim 1/g^2$,
the term proportional to $\zeta$ is the leading term on the
right-hand side of the gap equation (\ref{gapequation}); together with the
prefactor $g^2$ it is of order $O(\phi_0)$, and thus matches
the order of magnitude of the left-hand-side of the gap equation. 
The value of the coefficient $\zeta$ determines 
the constant $c$ in Eq.\ (\ref{gapsolution}).
This constant was first computed by Son \cite{son99}, 
\begin{equation}
\frac{c}{g} = \frac{\pi}{2 \bar{g}}\;, \; \; 
\bar{g} \equiv \frac{g}{3 \sqrt{2}} \;.
\label{c}
\end{equation}

The second and third terms in Eq.\ (\ref{gapequation}) contain subleading
contributions to the gap equation. These are 
proportional to a single power of the logarithm $\ln (\mu/\phi_0) \sim 1/g$.
For the term proportional to $\beta$, this logarithm is
the BCS logarithm from Eq.\ (\ref{BCSlog}).
In the other term, proportional to $\beta'$, this logarithm is explicitly 
present, but the BCS logarithm is absent, because the additional factor
$\epsilon_q^+$ cancels the one in the denominator of the integration
measure, ${\rm d}q/\epsilon_q^+$.
Together with the prefactor $g^2$, these terms constitute a contribution of
order $O(g \phi_0)$ to the right-hand-side of the gap equation,
{\it i.e.}, an $O(g)$ correction to the leading contribution.

The contribution proportional to $\beta$ arises from the
exchange of non-static magnetic and static electric
gluons \cite{pis00}. Both types of interactions are short-range:
they are screened on a distance scale $m_g^{-1}$, where $m_g$ is the
gluon mass; $m_g^2=N_f g^2 \mu ^2 /(6\pi^2)$, $N_f$ is the number of
quark flavors. Consequently, the collinear logarithm characteristic for
long-range interactions is absent, and one is left with 
the BCS logarithm. 
The contribution proportional to $\beta'$ arises from the
quark self-energy \cite{bro00,qwdhr}. It is parametrically
of the same order as the term multiplying $\beta$.
The coefficients $\beta$ and $\beta'$ in Eq.\ (\ref{gapequation})
determine the constant $b$ in Eq.\ (\ref{gapsolution}).
Son \cite{son99} was the first to give an estimate for the constant $b$, 
\begin{equation}
b = \frac{b_0}{g^5}\;,
\label{b}
\end{equation} 
with a constant $b_0$ of order $O(1)$, which could not be determined in the
approach of Ref.\ \cite{son99}.
In \cite{SchaferWilczek,pis00} the constant $b_0$ was
computed by solving the QCD gap equation including 
non-static magnetic and static electric gluon exchange, but
without taking into account the quark self-energy. In other words,
all terms $\sim \beta$ in Eq.\ (\ref{gapequation}) were collected,
but the term $\sim \beta'$ was neglected. The result is
\begin{equation}
\label{b0}
b_0 = 256\, \pi^4 \left( \frac{2}{N_f} \right)^{5/2}\ b_0'\;,
\end{equation}
with an undetermined constant $b_0'$ of order $O(1)$.
In \cite{SchaferWilczek,pis00}, {\it i.e.}, without
effects from the quark self-energy, $b_0'=1$. 
In \cite{bro00,qwdhr}, it was shown that the quark self-energy 
gives rise to a term $\sim \beta'$ in Eq.\ (\ref{gapequation}).
As this is parametrically also of subleading order in the
gap equation, it modifies the constant $b_0'$,
\begin{equation}
b_0' = \exp \left( - \frac{\pi^2 + 4}{8} \right) \simeq 0.177\; .
\label{b0'}
\end{equation}

The fourth term in Eq.\ (\ref{gapequation}) summarizes sub-subleading
contributions. These are of order $O(g^2 \phi_0)$.
It was argued in \cite{SchaferWilczek,pis00,ShusterRajagopal} that
at this order gauge-dependent terms enter the mean-field gap equation
for the color-superconducting gap parameter.
However, the gap parameter is in principle an observable quantity, 
and thus gauge independent. Therefore, it was concluded that one
has to go beyond the mean-field approach to
compute gauge-independent sub-subleading contributions to
the gap parameter.
It was also shown \cite{man1} that
effects from the finite lifetime of quasiparticles in the Fermi
sea influence the value of $\phi_0$ at this order. 
In weak coupling, the terms $\sim \alpha$ in Eq.\ (\ref{gapequation})
are suppressed by one power of $g$ compared to the subleading terms 
and therefore constitute an
order $O(g)$ correction to the prefactor $b$, as indicated in
Eq.\ (\ref{gapsolution}).

In this note, we first present the gap equation for the 
quasiparticle and quasi-antiparticle gap, including the
gauge-dependent
terms, and review previous arguments \cite{pis00}
on why the gauge dependence enters at sub-subleading order in the
gap equation. These arguments were actually incorrect, in that they
neglected additional powers of the gluon momentum 
in the gap equation.
Naively correcting for these powers, one obtains
that the gauge dependence enters already
at subleading order, giving rise to an extra prefactor 
$\sim \exp (3\, \xi_C/2)$ to the gap parameter (\ref{gapsolution}),
where $\xi_C$ is the gauge parameter for the gluon
propagator in a general Coulomb gauge. 
This result is similar to other claims made in the literature
\cite{miransky}. Finally, we demonstrate that
a careful calculation of the gauge-dependent term
on the correct quasiparticle mass shell
shows that the gauge dependence indeed enters only beyond subleading
order in the gap equation. 
Consequently, the gauge dependence does not affect
the $O(1)$ result for the prefactor of the gap, as was originally claimed.

Our units are $\hbar=c=k_B=1$, the metric tensor is $g^{\mu \nu} = 
{\rm diag}(+,-,-,-)$, and we work in a general Coulomb gauge, with
gauge parameter $\xi_C$. We denote 4-vectors with capital letters,
$K^\mu \equiv (k_0,{\bf k})$, $k \equiv |{\bf k}|$, 
$\hat{\bf k} \equiv {\bf k}/k$.

\section{GAUGE-DEPENDENT TERMS IN THE QCD GAP EQUATION}

For the sake of simplicity, let us focus on the condensation
of quarks with two massless flavors, forming Cooper pairs with 
total spin zero.
For the discussion of the gauge dependence, terms in the gap equation
arising from the quark self-energy \cite{qwdhr} can be
safely neglected.
Thus, the QCD gap equation reads, cf.\ Eq.\ (29) of \cite{pis00},
\begin{eqnarray}
\phi_h^e(K) & = & \frac{2}{3}\, g^2 \; T \sum_{q_0} \int 
\frac{{\rm d}^3{\bf q}}{(2\pi)^3}\;
\Delta_{\mu \nu}(K-Q) \, \left\{
\frac{\phi_h^e(Q)}{q_0^2 - (\epsilon^e_q)^2}  
    \, {\rm Tr} \left[{\cal P}_h^e({\bf k})\, \gamma^\mu \,
{\cal P}_{-h}^{-e}({\bf q})\, \gamma^\nu \right] \right. \nonumber \\
&   & \hspace*{5.5cm} \left.
+\, \frac{\phi_h^{-e}(Q)}{q_0^2- (\epsilon^{-e}_q)^2} 
   \, {\rm Tr} \left[{\cal P}_h^e({\bf k}) \, \gamma^\mu \,
{\cal P}_{-h}^e({\bf q}) \, \gamma^\nu \right] \right\} \,\, .
\label{QCDgapequation}
\end{eqnarray}
Here, $\phi_h^e(K)$ is the gap function for condensation of quarks
with chirality $h$ and energy $e$, $\epsilon_q^e \equiv
\sqrt{(\mu - e q)^2 + |\phi_h^e(Q)|^2}$, and 
\begin{equation}
{\cal P}_h^e({\bf k}) \equiv \frac{1+h \gamma_5}{2} \;
\frac{1+e \gamma_0 \mbox{\boldmath{$\gamma$}} \cdot \hat{\bf k}}{2} 
\end{equation}
are projectors \cite{pis99} onto states with chirality $h=\pm=r,\ell$ and
energy $e = \pm$. The Matsubara sum $\sum_{q_0}$ runs over fermionic
Matsubara frequencies $q_0 \equiv - i (2n+1) \pi T$. 
We first perform the Matsubara sum and then consider
the limit $T \rightarrow 0$.
The gluon propagator in a general Coulomb gauge is,
cf.\ Eq.\ (30) of \cite{pis00},
\begin{eqnarray}
\Delta_{00}(P) & = & \Delta_{l}(P) + \xi_C\, \frac{p_0^2}{p^4} \,\, , 
\nonumber  \\
\Delta_{0i}(P) & = & \xi_C\, \frac{p_0\, p_i}{p^4} \,\, , \label{Delta}\\
\Delta_{ij}(P) & = & (\delta_{ij} - \hat{p}_i \, \hat{p}_j)\,
\Delta_{t}(P) + \xi_C\, \frac{p_i\, p_j}{p^4} \,\, . \nonumber
\end{eqnarray}
To simplify the notation in the following, 
let us write the QCD gap equation (\ref{QCDgapequation}) in the form
\begin{equation}
\phi_h^e(K) = R_h^e(K) + \xi_C \, X_h^e(K)\,\, ,
\end{equation}
where $R_h^e(K)$ is the right-hand side of (\ref{QCDgapequation})
for $\xi_C = 0$, and introducing $P \equiv K - Q$, 
\begin{eqnarray}
X_h^e(K) & \equiv & \frac{2}{3}\, g^2 \; T \sum_{q_0} \int 
\frac{{\rm d}^3{\bf q}}{(2\pi)^3}\;
\frac{P_\mu \, P_\nu}{p^4} \, \left\{
\frac{\phi_h^e(Q)}{q_0^2 - (\epsilon^e_q)^2}  
    \, {\rm Tr} \left[{\cal P}_h^e({\bf k})\, \gamma^\mu \,
{\cal P}_{-h}^{-e}({\bf q})\, \gamma^\nu \right] \right. \nonumber \\
&   & \hspace*{4.3cm} \left.
+\, \frac{\phi_h^{-e}(Q)}{q_0^2- (\epsilon^{-e}_q)^2} 
   \, {\rm Tr} \left[{\cal P}_h^e({\bf k}) \, \gamma^\mu \,
{\cal P}_{-h}^e({\bf q}) \, \gamma^\nu \right] \right\} \,\, .
\label{X}
\end{eqnarray}
The traces can be readily evaluated,
\begin{eqnarray}
P_\mu\, P_\nu \,{\rm Tr} \left[{\cal P}_h^e({\bf k})\, \gamma^\mu \,
{\cal P}_{-h}^{-e}({\bf q})\, \gamma^\nu \right]
& = & \frac{1+\hat{\bf k} \cdot \hat{\bf q}}{2} \;
\left[p_0^2 - (k - q)^2 \right] \,\, ,\nonumber \\
P_\mu\, P_\nu \,{\rm Tr} \left[{\cal P}_h^e({\bf k})\, \gamma^\mu \,
{\cal P}_{-h}^e({\bf q})\, \gamma^\nu \right]
& = & \frac{1-\hat{\bf k} \cdot \hat{\bf q}}{2} \;
\left[p_0^2 - (k + q)^2 \right] \,\, .
\label{traces}
\end{eqnarray}
Since the gap equations for right- and left-handed gaps
decouple, we shall drop the index $h$ in the following.
We furthermore focus on the gap equation for the quasiparticle
gap function $\phi^+(K)$, {\it i.e.}, $e=+$ in Eqs.\
(\ref{QCDgapequation}) and (\ref{X}). 
In the next section, we argue that quasi-antiparticle
contributions to the gap equation for the quasiparticle gap are
negligible. We then proceed to estimate the magnitude of
the gauge-dependent terms in Section 4.

\section{NEGLECTING THE QUASI-ANTIPARTICLE MODES}

It is permissible to
neglect the contribution from the quasi-antiparticle excitations
$\sim \phi^-(Q)$ in the gap equation (\ref{QCDgapequation}), as has been
done in Eq.\ (\ref{gapequation}).
The reason is that the quasi-antiparticle gap
$\phi^-(K)$ is suppressed with respect to the quasiparticle gap
$\phi^+(K)$ by at least one power of $g$. 
This can be checked by simply power-counting
the contributions to the gap equation for $\phi^-(K)$ using
the arguments presented in the introduction. 
Taking the gap function on the quasi-antiparticle mass shell, 
$k_0 = \epsilon_k^-$, {\it i.e.}, 
$\phi^-(K) \equiv \phi^-(\epsilon_k^-,{\bf k}) \equiv \phi_k^-$,
and performing the Matsubara sum over $q_0$ (which puts the
internal gap functions on the mass shell, $\phi^\pm (Q) \equiv
\phi^\pm(\epsilon_q^\pm,{\bf q}) \equiv \phi_q^\pm$) and the 
integral over the angle between external quark three-momentum ${\bf k}$
and internal quark three-momentum ${\bf q}$, 
this gap equation has a similar structure as Eq.\ (\ref{gapequation}), with
the obvious replacements $\phi^+_k \rightarrow \phi^-_k,\,
\epsilon_k^+ \rightarrow \epsilon_k^-$.
(The contribution from quasi-antiparticles to this
gap equation, which is still present in Eq.\ (\ref{QCDgapequation}),
can be discarded, as it does not even have a BCS logarithm.)
Since $\epsilon_k^- \simeq \mu + k$, 
the collinear logarithm from almost static,
magnetic gluon exchange in Eq.\ (\ref{gapequation})
is only of order $O(1)$. This leaves the BCS logarithm.
At the Fermi surface
\begin{equation}
\phi^-_\mu \sim g^2 \, \ln \left(\frac{\mu}{\phi_0} \right) \, \phi_0
\,\, .
\end{equation}
With Eq.\ (\ref{gapsolution}) one obtains $\phi^-_\mu \sim g\, \phi_0$,
which proves our assertion. 

The complete contribution from
quasi-antiparticle excitations to the equation for the
quasiparticle gap is suppressed relative to the quasiparticle contribution
even more than just by a single power of $g$.
Neglecting a possible angular dependence entering through the traces in
Eq.\ (\ref{QCDgapequation}) and performing the Matsubara sum
which picks up the poles at $q_0 = \pm \epsilon_q^- \simeq \pm (\mu+q)$,
the quasi-antiparticle contribution is proportional to
\begin{equation}
\frac{\phi^-_q}{\epsilon_q^-} 
\sim \frac{g\, \phi_q^+}{\epsilon_q^-} 
\simeq \frac{g\, \epsilon_q^+}{\mu + q} \, 
\frac{\phi^+_q}{\epsilon_q^+} \,\, ,
\label{comparison}
\end{equation}
{\it i.e.}, at the Fermi surface 
it is suppressed by a factor $\sim g \phi_0/(2\mu)$ relative
to the quasiparticle contribution (which is proportional to the last
factor in Eq.\ (\ref{comparison})).

\section{THE GAP IS GAUGE DEPENDENT OFF THE MASS SHELL ...}

The argument of \cite{pis00}, as to why the gauge-dependent
terms do not enter into the gap equation at leading and subleading order,
starts with neglecting the terms proportional to $p_0^2$
in Eq.\ (\ref{traces}). This is because on the mass shell such terms
are of order $p_0^2 \equiv (k_0 - q_0)^2 \sim 
(\epsilon_k^+ \pm \epsilon_q^+)^2 \sim \phi_0^2$.
One is then left with the spatially longitudinal terms, cf.\
Eq.\ (134) of \cite{pis00},
\begin{equation}
{\rm Tr} \left[{\cal P}_h^+({\bf k})\, 
\mbox{\boldmath{$\gamma$}} \cdot \hat{\bf p} \,
{\cal P}_{-h}^{-}({\bf q})\, 
\mbox{\boldmath{$\gamma$}} \cdot \hat{\bf p} \right] 
\sim - \frac{1+ \hat{\bf k} \cdot \hat{\bf q}}{2} \,
\frac{(k-q)^2}{p^2} \,\, .
\label{gaugedependence1}
\end{equation}
As $k$ and $q$ approach the Fermi surface, this term vanishes, completing
the argument that gauge-dependent terms do not enter at leading
and subleading order in the gap equation.

This is, however, a very general argument. So let us explicitly
compute $X^+(K)$ from Eq.\ (\ref{X}) with (\ref{gaugedependence1}),
just to make sure. We had made 
the seemingly innocuous assumption 
of neglecting terms $\sim p_0^2$. In doing so, implicitly we will
be working off the mass shell, and so find 
that the gap is gauge dependent. In the next section, we show that
this goes away on the mass shell.

Under the present approximations, there is no term depending on 
$q_0$ except for the denominator in (\ref{X}). Thus, one can 
immediately perform the Matsubara sum over $q_0$ using
\begin{equation}
T \sum_{q_0} \frac{\phi^+(Q)}{q_0^2 - (\epsilon_q^+)^2}
\equiv -\frac{\phi_q^+}{2 \epsilon_q^+}\,
\left[1-2\,n_F\left(\frac{\epsilon_q^+}{T}\right) \right] \,\, ,
\end{equation}
where $n_F(x) \equiv (e^x +1)$ is the Fermi distribution.
For $T=0$, $n_F(\epsilon_q^+/T)$ vanishes, as $\epsilon_q^+ \geq
\phi_q^+ >0$. Accounting
for an additional factor $1/p^2$ from Eq.\ (\ref{X}),
one then integrates over ${\rm d}^3 {\bf q}$.
The integrand does not depend on the polar angle; therefore
${\rm d}^3 {\bf q} \equiv 2 \pi \; {\rm d}q \, q^2\; {\rm d} \cos \theta$. 
Here, $\cos \theta \equiv \hat{\bf k} \cdot \hat{\bf q}$. Substituting 
$p \equiv |{\bf k} - {\bf q}| $ for $\cos \theta$, 
the following integral appears in Eq.\ (\ref{X}),
\begin{equation}
\int_{|k-q|}^{k+q} {\rm d} p \, p \,
\frac{(k+q)^2 - p^2}{p^4} =  \frac{2 k q}{(k - q)^2} - 
\ln \frac{k+q}{|k-q|} \,\, .
\end{equation}
As this term is multiplied by $(k-q)^2$ on account of
(\ref{gaugedependence1}), the singularity at $k=q$ is 
rendered harmless.
Inserting this into Eq.\ (\ref{X}), one obtains
\begin{equation}
X^+ (k)  = \frac{g^2}{24\, \pi^2} 
\int {\rm d} q\, \frac{\phi_q^+}{\epsilon_q^+}
\, \left[ \frac{q}{k} - \frac{(k-q)^2}{2k^2}
\ln \frac{k+q}{|k-q|} \right] \,\, .
\label{gaugedependence2}
\end{equation}
The momentum dependence of the gap function restricts the $q$-integration
to a narrow interval around the Fermi surface, $\mu - \delta \leq q \leq
\mu + \delta, \, \delta \ll \mu$ \cite{pis00}.
At the Fermi surface, $k = \mu$, and introducing $\xi \equiv q - \mu$
one obtains
\begin{equation}
X^+ (\mu) = \frac{g^2}{12 \, \pi^2}
\int_0^\delta {\rm d} \xi  \, \frac{\phi^+_\xi}{\epsilon_\xi}\, 
\left( 1 - \frac{\xi^2}{4\mu^2}\, \ln \frac{4 \mu^2}{\xi^2} \right) \,\, ,
\label{Ximu}
\end{equation}
where we have neglected some 
terms of order $\xi/\mu \leq \delta/\mu \ll 1$, and
where $\epsilon_\xi \equiv \sqrt{\xi^2 + |\phi_\xi^+|^2}$.
As one approaches the Fermi surface, $\xi \rightarrow 0$,
the second term in parentheses vanishes. 
Consequently, the first term is the dominant one, and it
gives rise to a BCS logarithm,
\begin{equation}
X^+ (\mu) = \frac{g^2}{12\, \pi^2} 
\int_0^\delta {\rm d} \xi \, \frac{\phi^+_\xi}{\epsilon_\xi}
\sim g^2 \phi_0 \ln \left( \frac{\mu}{\phi_0} \right) \,\, .
\label{Ximu2}
\end{equation}
Such a term is of the same order as terms $\sim \beta$ in 
Eq.\ (\ref{gapequation}), {\it i.e.}, of {\em subleading\/} order
in the gap equation~! 
It thus affects the prefactor $b$ in Eq.\ (\ref{gapsolution}).
A careful analysis shows that $b$ is multiplied by a factor
$\exp (3 \, \xi_C/2)$.
In pure Coulomb gauge, $\xi_C = 0$, this factor
is unity, and one obtains the previous result for $b$, Eqs.\ (\ref{b})
and (\ref{b0}).
A similar dependence of $b$ on the gauge parameter was
also reported for covariant gauges \cite{miransky}.
The appearance of a gauge dependence in those calculations is, however, 
not surprising, as the gap parameter is not calculated on the 
quasiparticle mass shell, but for Euclidean (imaginary) energies.
Away from the quasiparticle mass shell, the gap is not an
observable quantity, and therefore may depend on the choice of gauge.

The above argument is, however, still incorrect, as terms
proportional to $p_0^2$ were neglected. Therefore, as mentioned
previously, we are also computing off the mass shell. Restoring these
terms and putting $k_0$ on the quasiparticle mass shell,
we shall see that the gauge dependence enters neither at leading
nor subleading order in the gap equation.

\section{... AND INDEPENDENT OF GAUGE ON THE MASS SHELL}

We now compute the gap on the correct mass shell. To this end,
the seemingly innocuous terms $\sim p_0^2$ in Eq.\ (\ref{traces})
must not be neglected, since $p_0 = k_0 - q_0$.
To perform the Matsubara sum over these
terms, one uses the following trick:
\begin{equation}
T \sum_{q_0} p_0^2\, \frac{\phi^+(Q)}{q_0^2 - (\epsilon_q^+)^2}
\equiv \lim_{\tau \rightarrow 0} \; (k_0 + \partial_\tau)^2\;
T \sum_{q_0} e^{-q_0 \tau}\, \frac{\phi^+(Q)}{q_0^2 - (\epsilon_q^+)^2}\,\,.
\end{equation}
The Matsubara sum can be computed in the standard way. Taking 
$T \rightarrow 0$, the final result is
\begin{equation}
\lim_{T \rightarrow 0}
T \sum_{q_0} p_0^2\, \frac{\phi^+(Q)}{q_0^2 - (\epsilon_q^+)^2}
= -\frac{\phi_q^+}{2 \epsilon_q^+}\,(k_0 - \epsilon_q^+)^2 \,\, .
\end{equation}
Putting everything together and taking the external energy $k_0$ on the
mass shell, $k_0 = \epsilon_k^+$, one obtains for $X^+_k \equiv
X^+(\epsilon_k^+,{\bf k})$ the following equation:
\begin{equation}
X_k^+  = \frac{g^2}{48\, \pi^2 k^2} \int {\rm d} q \;
\frac{\phi_q^+}{\epsilon_q^+} \,
\left[\frac{2 k q}{(k - q)^2} - \ln \frac{k+q}{|k-q|}  \right]
\; \left[ (k-q)^2 - (\epsilon_k^+ - \epsilon_q^+)^2 \right]  \,\,.
\end{equation}
We again introduce the variable $\xi \equiv q - \mu$ and
neglect terms $\sim \xi/\mu \ll 1$.
At the Fermi surface, $k = \mu$, we are then left with
\begin{equation}
X_\mu^+  =  
\frac{g^2}{12\, \pi^2} \int_0^\delta {\rm d} \xi\,
\frac{\phi_\xi^+}{\epsilon_\xi} \,
\left( 1 - \frac{\xi^2}{4  \mu^2} \, \ln \frac{4 \mu^2}{\xi^2}  \right)
\,\left[ 1 - \frac{(\phi_0 - \epsilon_\xi)^2}{\xi^2}  \right]\,\, .
\label{result}
\end{equation}
As in Eq.\ (\ref{Ximu}), the logarithm in parentheses is completely
innocuous because of the prefactor $\xi^2$, 
and can be neglected in the following.

The difference to Eq.\ (\ref{Ximu}) is 
the second term in brackets, which arises from the
$p_0^2$ term in Eq.\ (\ref{traces}). Expanding this term 
around the Fermi surface ($\xi = 0$),
one realizes that it is of order $\xi^2$, and therefore negligible
compared to the first term in brackets.
Naively, one would now conclude that 
there is a large BCS logarithm $\int {\rm d}\xi/\epsilon_\xi
\sim \ln (\mu/\phi_0)$, just like in Eq.\ (\ref{Ximu2}),
and the gauge-dependent term contributes to
subleading order in the gap equation.
This is, however, incorrect. The BCS logarithm ``builds up'' 
as $\xi$ approaches the Fermi surface at $\xi = 0$. This build-up
requires a coefficient of order $O(1)$ over the whole range of
integration.

In contrast to Eq.\ (\ref{Ximu}), this is not the case here, because
the factor $1$ in Eq.\ (\ref{Ximu}) is replaced by
$1 -(\phi_0 - \epsilon_\xi)^2/\xi^2$ in Eq.\ (\ref{result}).
To see the effect on the magnitude of the gauge-dependent
terms in the gap equation, split the integral over $\xi$
into two parts, one from $0$ to $\kappa \phi_0$,
with $\kappa \gg 1$, and one from $\kappa \phi_0$ to $\delta$.
In the first integral
\begin{equation}
1 - \frac{(\phi_0 - \epsilon_\xi)^2}{\xi^2}
\sim O(1)\,\, .
\end{equation}
In the second integral,
\begin{equation}
1 - \frac{ (\phi_0 - \epsilon_\xi)^2}{ \xi^2}
\sim  \frac{2\, \phi_0 \epsilon_\xi}{\xi^2}\,\, .
\label{2ndint}
\end{equation}
Inserting this into (\ref{result}), one obtains 
the order-of-magnitude estimate 
\begin{equation}
X_\mu^+ \sim g^2\, \phi_0 \; \left[ \int_0^{\kappa \phi_0}
\frac{{\rm d} \xi}{\epsilon_\xi} 
+ \phi_0 \int_{\kappa \phi_0}^\delta \frac{{\rm d} \xi}{\xi^2}
\right] \simeq g^2 \, \phi_0 \; 
\left[ \ln (2 \kappa) - \frac{\phi_0}{\delta} + \frac{1}{\kappa}
\right] \sim g^2 \, \phi_0\,\,.
\label{finalresult}
\end{equation}
In the first integral, the integration measure would in principle give
rise to a BCS logarithm, if the upper limit of integration was large,
$\sim \delta$, and not small, $\sim \phi_0$. 
In the second integral, the factor $\epsilon_\xi$ 
in the numerator of (\ref{2ndint})
cancels with the one in the integration measure 
${\rm d}\xi/\epsilon_\xi$ and thus prevents the ``build-up'' of the
BCS logarithm.

Comparing the final result (\ref{finalresult})
with the discussion in the introduction,
we conclude that the gauge-dependent term is obviously of
sub-subleading order in the gap equation (\ref{gapequation}).
In other words, the mean-field gap equation for the
color-superconducting gap in QCD is gauge independent to leading
and subleading order. Consequently, the gauge-dependent terms 
influence the prefactor of the gap in Eq.\ (\ref{gapsolution})
only at order $O(g)$. The present note thus confirms by analytical
means the numerical results of Ref.\ \cite{ShusterRajagopal}.

\section*{ACKNOWLEDGEMENTS}

D.H.R.\ thanks I.\ Shovkovy and Q.\ Wang for discussions.

\end{document}